\documentclass[showpacs,preprintnumbers,superscriptaddress]{revtex4}
\usepackage{CJK}
\usepackage{amsmath,amssymb,graphicx,bm}
\begin{document}

\title{Quantum gravity corrections to accretion onto a Schwarzschild black hole}

\author{Rongjia Yang \footnote{Corresponding author}}
\email{yangrongjia@tsinghua.org.cn}
\affiliation{College of Physical Science and Technology, Hebei University, Baoding 071002, China}
\affiliation{Hebei Key Lab of Optic-Electronic Information and Materials,Hebei University, Baoding 071002, China}
\affiliation{State Key Laboratory of Theoretical Physics, Institute of Theoretical Physics, Chinese Academy of Sciences,
Beijing 100190, China}

\begin{abstract}
Quantum gravity corrections to accretion onto a Schwarzschild black hole are considered in the context of asymptotically safe scenario. The possible positions of the critical points are discussed and the general conditions for critical points are obtained. The explicit expressions for matter density compression and temperature profile both below the critical radius and at the event horizon are derived. For polytropic matter, we determine the corrected temperature and the integrated flux resulting from quantum gravity effects at the event horizon, which might be as a test of asymptotically safe scenario.
\end{abstract}


\maketitle

\section{Introduction}
Accretion of matter onto astronomical objects is an important phenomenon of long-standing interest to astrophysicists and the most likely scenario to explain the high energy output from quasars and active galactic nuclei. In the context of Newtonian gravity, the stationary, spherically symmetric and transonic accretion of adiabatic fluids onto astrophysical objects was investigated in \citep{Bondi:1952ni}, which generalizes the earlier results which discussed pressure-free gas being dragged onto a massive central object \citep{hoyle1939effect,Bondi:1944jm}. Michel considered the steady-state spherical symmetric flow of matter into or out of a condensed object in the framework of general relativity \citep{michel1972accretion}. Based on Michel's work, the problem of the critical points of accretion was examined in \citep{begelman1978accretion}. Accretion onto a moving black hole was studied in \citep{petrich1988accretion}, onto a charged black hole was investigated in \citep{michel1972accretion,Jamil:2008bc}, onto a higher dimensional black hole was analyzed in \citep{Giddings:2008gr,Sharif:2011ih,John:2013bqa,Debnath:2015yva}, onto a black hole in a string cloud background was considered in \citep{Ganguly:2014cqa}, onto a Kerr-Newman black hole was dealt with in \citep{Babichev:2008dy, JimenezMadrid:2005rk,Bhadra:2011me}, and onto Schwarzschild-(anti-)de Sitter spacetimes or onto cosmological black holes was examined in \citep{Mach:2013fsa, Mach:2013gia, Karkowski:2012vt}. The case for general relativistic spherical accretion with and without back-reaction were discussed in \citep{Malec:1999dd, Karkowski:2005zn, Mach:2008wq, Dokuchaev:2011gt, Babichev:2012sg}. It was shown that accretion of phantom energy onto a black hole induces a gradual decrease of the black hole mass \citep{Babichev:2004yx, Babichev:2005py}. It was found, however, that the physical black hole mass may instead increase due to the accretion of phantom energy if using solutions describing black holes in a background Friedmann-Robertson-Walker universe \citep{Gao:2008jv}.

All the above researches based on the classical theories of gravity which are well known to be perturbatively nonrenormalizable. However, Weinberg postulated that the effective quantum field theory of gravity might exhibits asymptotic safety \citep{weinberg1979general}, a non-Gaussian fixed point of the renormalization group flow of the couplings of the gravitational theory which controls the behavior of the theory at very high energies and ensures the absence of unphysical ultraviolet divergences. This interesting idea has been applied to investigate the existence of the UV fixed point in various theories, such as Einstein gravity, scalar-tensor theory, and $f(R)$ gravity (for reviews see \citep{Litim:2006dx, Niedermaier:2006wt, percacci2009approaches}, and references therein). It has been suggested that there is also the possibility of an infrared fixed point in quantum gravity, which has been explored in cosmology \citep{Bonanno:2001xi,Hindmarsh:2011hx,Ahn:2011qt,Yang:2012tm}. However, the possibility of an IR fixed point in quantum gravity is still a speculative idea which has yet no direct evidence from any computed $\beta$-functions, which is worthy of further studying.

In this paper, we will analyze quantum gravity corrections to accretion onto black holes in the context of asymptotically safe gravity. We will consider steady, spherical accretion onto a static and spherically symmetric black hole. We will analytically determine the critical point, critical fluid velocity, temperature, the mass accretion rate, and subsequently observed total integrated flux.

The paper is outlined as follows. In next section, we will present the quantum gravity corrections to accretion onto black holes and obtain the basic equations for accretion. In Sec. III, we will determine the critical points and conditions of the accretion. In Sec. IV, we will discuss the polytropic solution and determine the corresponding quantities.  Finally, we will briefly summarize and discuss our results in section V.

\section{Fundamental equations}
We consider a static and spherically symmetric Schwarzschild black hole. Assuming that the leading quantum gravity corrections to Schwarzschild black hole metrics are accounted for by replacing Newton's coupling constant through a `running' coupling which evolves under the renormalisation group equations for gravity, the renormalisation group improved Schwarzschild solution can be written as \citep{Bonanno:2000ep}
\begin{eqnarray}
\label{Sp}
ds^2=\left[1-\frac{2MG(r)}{r}\right]dt^2-\left[1-\frac{2MG(r)}{r}\right]^{-1}dr^2-r^2\Big(d\theta^2+\sin^2\theta d\varphi^2\Big),
\end{eqnarray}
where
\begin{eqnarray}
\label{gr}
G(r)=\frac{G_0r^3}{r^3+\bar{\omega}(r+\gamma G_0M)},
\end{eqnarray}
where $\bar{\omega}=\tilde{\omega} G_0(=\tilde{\omega} G_0\hbar/c^3)$, $G_0$ is Newton's constant, $M$ is the mass of the black hole measured by an observer at infinity, $\gamma$ and $\tilde{\omega}$ are constants coming from an appropriate cutoff identification and from the non-perturbative renormalization group theory, respectively. At large distances, the leading correction to Newton's constant is given by $G(r)=G_0-\bar{\omega}G_0/r^2+O(1/r^3)$. At small distances, it behaves like $G(r)=r^3/(\gamma\bar{\omega}M)+O(r^4)$. Since the qualitative properties of the solution (\ref{Sp}) are insensitive to the precise value of $\gamma$ for $r\gg G_0M$, it is usual to choose $\gamma=0$ \citep{Bonanno:2000ep}. Comparison with the standard perturbative quantization of Einstein's gravity, the precise value of $\tilde{\omega}$ is found to be $\tilde{\omega}=167/(30\pi)$ \citep{BjerrumBohr:2002kt}. The properties of the solution (\ref{Sp}) also do not depend on the precise value of $\omega$ as long as it is strictly positive.

The horizons of the improved solution (\ref{Sp}) can be obtained by solving $g_{00}=0$ and are found to be depend on whether the mass of the black hole is bigger, equal or smaller than a critical value $M_{\rm{cr}}$. For $\gamma=0$, the critical mass is $M_{\rm{c}}=\sqrt{\bar{\omega}}/G_0\simeq 1.33 m_{\rm{p}}$ with $m_{\rm{p}}$ the Planck mass. If $M>M_{\rm{c}}$, there are two positive real solutions: $r_{\rm{h}\pm}=MG_0[1\pm\sqrt{1-\bar{\omega}/(G_0M)^2}]$ satisfying $r_{\rm{h+}}>r_{\rm{h-}}$ for $g_{00}=0$. The outer solution $r_{\rm{h+}}$ is the Schwarzschild horizon with quantum corrections taken into account and can be expanded as
\begin{eqnarray}
\label{rp}
r_{\rm{h+}}=2G_0M\left[1-\frac{1}{4}\frac{\bar{\omega}}{(G_0M)^2}\right]=2G_0M\left[1-\frac{1}{4}\tilde{\omega}\left(\frac{m_{\rm{p}}}{M}\right)^2\right],
\end{eqnarray}
if the mass of black hole is much bigger than Planck's mass. There is only one horizon, $r_{\rm{h}}=MG_0$, if $M=M_{\rm{c}}$, and there is no horizon if $M<M_{\rm{c}}$.

We consider the steady-state radial inflow of gas onto a black hole described by the improved Schwarzschild metric (\ref{Sp}). The gas is approximated as a perfect fluid specified by the stress energy tensor
\begin{eqnarray}
\label{em}
T_{\mu\nu}=(\mu+p)u_{\mu}u_{\nu}-g_{\mu\nu}p,
\end{eqnarray}
where $\mu$ and $p$ are the total proper energy density and the proper pressure of the fluid correspondingly, $u^{\mu}=dx^{\mu}/ds$ is the fluid 4-velocity with $u^{\mu}u_{\mu}=1$. We denote the radial component of the 4-velocity as $u=dr/ds$. Since the components of velocity for $\mu>1$ vanish, it is easy to get $u^0=[1-2MG(r)/r+u^2]^{1/2}/[1-2MG(r)/r]$. Generally, the total energy density is given by $\mu=\rho+\varepsilon$, where $\rho$ is the proper matter density and $\varepsilon$ is the internal energy. The relation between $\rho$ and $\mu$ is, $d\mu/d\rho=(\mu+p)/\rho+T\rho(ds/d\rho)$, where $T$ is the temperature and $s$ is the specific entropy \citep{shapiro2008black, Naskar:2007su}. An important physical quantity in the description of the accretion is the speed of sound, which under isentropic conditions is expressed in
terms of the thermodynamic quantities \citep{shapiro2008black, Naskar:2007su}
\begin{eqnarray}
\label{sop}
c^2_{\rm s}=\left(\frac{\partial p}{\partial \mu}\right)_s=\frac{\mu+p}{\rho}.
\end{eqnarray}

The conservation of the mass flux $J^{\mu}=\rho u^{\mu}$ gives
\begin{eqnarray}
\label{con1}
J^{\mu}_{~;\mu}=(\rho u^{\mu})_{;\mu}=0,
\end{eqnarray}
where $;$ denotes the covariant derivative. For the improved Schwarzschild black hole (\ref{Sp}), Eq. (\ref{con1}) can be reformulated as
\begin{eqnarray}
\label{con1s}
\frac{d}{dr}(\rho u r^2)=0,
\end{eqnarray}
which for a perfect fluid gives the integration as
\begin{eqnarray}
\label{con1s1}
\rho u r^2=C_1,
\end{eqnarray}
where $C_1$ is a constant of integration. Integrating equation (\ref{con1s}) over the spacial volume, we obtain
\begin{eqnarray}
\label{acc}
\dot{M}=4\pi r^2u\rho,
\end{eqnarray}
where $\dot{M}$ is an integral constant with dimensions of mass per unit time. In fact, this is the improved Bondi's mass accretion rate. The mass of a black hole is a dynamic quantity in the context of astrophysics. With the improved Bondi's mass accretion rate, we can get the improved observed total integrated flux
\begin{eqnarray}
\label{intflux}
F_\nu=\frac{L_\nu}{4\pi d^2_{\rm {L}}},
\end{eqnarray}
where $L_\nu=\epsilon_\nu \dot{M}$ is the surface luminosity measured at infinity with $\epsilon_\nu$ a constant and $d_{\rm {L}}$ is the luminosity distance which can be measured. The total flux contribution from all the black hole in the observable universe are obtained by summing over (\ref{intflux}).

Assuming that the in-falling fluid is globally spherically symmetric, the energy-momentum conservation $T^{\mu}_{~\nu;\mu}=0$ gives
\begin{eqnarray}
\label{con2s}
\frac{d}{dr}\left[(\mu+p)u r^2\left(1-\frac{2MG(r)}{r}+u^2\right)^{1/2}\right]=0,
\end{eqnarray}
for the $\nu=0$ component, which implies that
\begin{eqnarray}
\label{con2s1}
r^2(\mu+p)u\left[1-\frac{2MG(r)}{r}+u^2\right]^{1/2}=C_2,
\end{eqnarray}
where $C_2$ is a constant of integration. For $\nu=1$, the energy-momentum conservation $T^{\mu}_{~\nu;\mu}=0$ gives
\begin{eqnarray}
\label{con11}
u\frac{du}{dr}=-\frac{1-\frac{2MG(r)}{r}+u^2}{\mu+p}\frac{dp}{dr}-\frac{MG(r)}{r^2}+\frac{MG'(r)}{r}.
\end{eqnarray}
Eqs. (\ref{con1s1}), (\ref{acc}), and (\ref{con2s1}) or (\ref{con11}) are fundamental conservation equations for the flow of matter onto the improved Schwarzschild black hole (\ref{Sp}) where the back-reaction of matter is ignored. Dividing and then squaring Eqs. (\ref{con1s1}) and (\ref{con2s1}), we derive improved Bernoulli equation
\begin{eqnarray}
\label{con3}
\left(\frac{\mu+p}{\rho}\right)^2\left[1-\frac{2MG(r)}{r}+u^2\right]=\left(\frac{\mu_\infty+p_\infty}{\rho_\infty}\right)^2.
\end{eqnarray}
Differentiating of Eqs. (\ref{con1s1}) and (\ref{con3}) and eliminating $d\rho$, we obtain
\begin{eqnarray}
\label{cce}
\frac{du}{u}\left(V^2-\frac{u^2}{W^2}\right)+\frac{dr}{r}\left[2V^2-\frac{M}{W^2}\left(\frac{G(r)}{r}-G'(r)\right)\right]=0,
\end{eqnarray}
or
\begin{eqnarray}
\label{cce1}
\frac{du}{dr}=-\frac{u}{r}\frac{2V^2-\frac{M}{W^2}\left[\frac{G(r)}{r}-G'(r)\right]}{V^2-\frac{u^2}{W^2}},
\end{eqnarray}
where
\begin{eqnarray}
G'(r)\equiv \frac{dG(r)}{dr}=\frac {2G_0r}{\bar{\omega}+r^{2}}-\frac {2G_0r^{3}}{\left(\bar{\omega}+r^{2} \right) ^{2}},
\end{eqnarray}
\begin{eqnarray}
\label{w}
W^2=1-\frac{2MG(r)}{r}+u^2,
\end{eqnarray}
and
\begin{eqnarray}
\label{V}
V^2=\frac{d\ln(\mu+p)}{d\ln \rho}-1.
\end{eqnarray}
It is easy to see that $V^2$ is just equal to the sound speed, $c^2_{\rm s}=V^2$.

If no effects of quantum gravity are taken into accounted, Eqs. (\ref{con1s1}), (\ref{acc}), (\ref{con2s1}), (\ref{con11}), and (\ref{cce}) or (\ref{cce1}) reduce to the basic equations for steady spherically symmetric accretion onto Schwarzschild black hole \citep{michel1972accretion}.

\section{Critical accretion}
If at any point the denominator of the right hand side of the equation (\ref{cce1}) vanishes, the corresponding numerator must also vanish at that point. This point is called the critical point or sonic point of the flow \citep{Bondi:1952ni, Malec:1999dd, Karkowski:2012vt}. Only solutions passing through critical points correspond to material falling or flowing out of the object with monotonically increasing velocity along the particle trajectory, meaning that the flow is smooth at all points of spacetime. Setting the denominator and the numerator on the right hand side of the equation (\ref{cce1}) to zero, we get the critical point conditions:
\begin{eqnarray}
\label{critic}
u^2_*=\frac{M}{2}\left[\frac{G(r_*)}{r_*}-G'(r_*)\right]=\frac{G_0Mr_*(r^2_*-\bar{\omega})}{2(r^2_*+\bar{\omega})^2},
\end{eqnarray}
and
\begin{eqnarray}
\label{critic1}
V^2_*=\frac{u^2_*}{W^2(r_*)}={\frac {G_0Mr_* \left(\bar{\omega}-r^{2}_* \right)}{5G_0M\bar{\omega} r_*+3G_0Mr^{3}_*-2\bar{\omega}^{2}-4\bar{\omega} r^{2}_*-2r^{4}_*}},
\end{eqnarray}
where the subscript ``*" denote values taken at the critical point. Note that we can retrieve the results for accretion of the fluid onto a Schwarzschild black hole by setting $\bar{\omega}=0$ in the above equations \citep{michel1972accretion, Malec:1999dd}. Close to the compact object the failing fluid can exhibit various behaviors near the critical point of accretion.

A physically acceptable solution of equation (\ref{cce1}) exists if $u^2_*\geq 0$ and $V^2_*\geq 0$; hence one can easily get
\begin{eqnarray}
\label{ineq}
r^2_*-\bar{\omega}\geq0,
\end{eqnarray}
from $u^2_*\geq 0$ and
\begin{eqnarray}
\label{ineq1}
\left\{ \begin{array}{l@{\quad \quad } l}
\displaystyle r^2_*-\bar{\omega}\geq0,\\
\displaystyle -2r^{4}_*+3G_0Mr^{3}_*+5G_0M\bar{\omega} r_*-2\bar{\omega}^{2}-4\bar{\omega} r^{2}_*\leq 0,
\end{array}
\right.
\end{eqnarray}
or
\begin{eqnarray}
\label{ineq2}
\left\{ \begin{array}{l@{\quad \quad } l}
\displaystyle r^2_*-\bar{\omega}\leq0,\\
\displaystyle  -2r^{4}_*+3G_0Mr^{3}_*+5G_0M\bar{\omega} r_*-2\bar{\omega}^{2}-4\bar{\omega} r^{2}_*\geq 0,
\end{array}
\right.
\end{eqnarray}
from $V^2_*\geq 0$. Equation (\ref{ineq}) gives $r_*\geq\sqrt{\bar{\omega}}$. Obviously the quantum gravity effects change characteristics of the critical points. Because equations (\ref{ineq2}) contradicts with (\ref{ineq}), we only need to take into account equations (\ref{ineq1}) from which we get roots physically representing the locations of the critical or sonic points of the flow near the black hole. If we do not consider the quantum gravity effects, equations (\ref{ineq1}) reduce to the results in \citep{michel1972accretion, Malec:1999dd}. Like quantum gravity effects, back-reaction of the fluid can also change characteristics of the critical point, for details see \citep{Malec:1999dd}.

Equations (\ref{ineq1}) is so complicated that we cannot obtain the solutions accurately. Though we can not solve the equations (\ref{ineq1}) analytically, we can roughly know where the roots will be. We firstly discuss equation $-2r^{4}_*+3G_0Mr^{3}_*\leq 0$. It is easy to find it holds for $r_*\geq 3MG_0/2$. Secondly, we consider equation $5G_0M\bar{\omega} r_*-2\bar{\omega}^{2}-4\bar{\omega} r^{2}_*\leq 0$. We find that if $M\geq 4\sqrt{2}\sqrt{\bar{\omega}}/(5G_0)$ it holds for $r_*\geq 5MG_0[1+\sqrt{1-32\bar{\omega}/(5MG_0)^2}/]8$ or $0\leq r_*\leq 5MG_0[1-\sqrt{1-32\bar{\omega}/(5MG_0)^2}]/8$. If $M<4\sqrt{2}\sqrt{\bar{\omega}}/(5G_0)$ it holds for $r_*\geq 0$.

So if $4\sqrt{2}\sqrt{\bar{\omega}}/(5G_0)<M\leq 6\sqrt{\bar{\omega}}/(5G_0)$, we can conclude that there exists $r_1$ to make equations (\ref{ineq1}) holds if $r_*\geq r_1$, where $r_1$ satisfies $\sqrt{\bar{\omega}} \leq r_1 < 3MG_0/2$, $-2r^{4}_1+3G_0Mr^{3}_1+5G_0M\bar{\omega} r_1-2\bar{\omega}^{2}-4\bar{\omega} r^{2}_1= 0$, and very close to $3MG_0/2$.

If $M> 6\sqrt{\bar{\omega}}/(5G_0)$, $r_1$ satisfies $5MG_0[1+\sqrt{1-32\bar{\omega}/(5MG_0)^2}]/8< r_1 < 3MG_0/2$, $-2r^{4}_1+3G_0Mr^{3}_1+5G_0M\bar{\omega} r_1-2\bar{\omega}^{2}-4\bar{\omega} r^{2}_1= 0$, and very close to $3MG_0/2$.

If $M< 4\sqrt{2}\sqrt{\bar{\omega}}/(5G_0)$, there also exists $r_2$ to make equations (\ref{ineq1}) holds if $r_*\geq r_2$, where $r_2$ satisfies $\sqrt{\bar{\omega}}\leq r_2< 3MG_0/2$, $-2r^{4}_2+3G_0Mr^{3}_2+5G_0M\bar{\omega} r_2-2\bar{\omega}^{2}-4\bar{\omega} r^{2}_2= 0$, and close to $3MG_0/2$.

Now we can discuss the critical behaviors of the fluid accreting on to the improved Schwarzschild hole. (a) If $M> M_{\rm{c}}\equiv\sqrt{\bar{\omega}}/G_0$, there are two horizons: $r_{\rm{h}\pm}=MG_0[1\pm\sqrt{1-\bar{\omega}/(G_0M)^2}]$, and there exists $r_*$ satisfying equations (\ref{ineq1}) and $r_{\rm{h}+}>r_*>r_{\rm{h}-}$ or $r_*>r_{\rm{h}+}$; in other words, the critical points may be outside the outer horizon or may be between the outer and the inter horizon, see for example, for $M=2\sqrt{3}\sqrt{\bar{\omega}}/(3G_0)$, we get $r_{\rm{h}+}=3MG_0/2$, in this case $r_*$ can be larger than $r_{\rm{h}+}$ or less than $r_{\rm{h}+}$. (b) If $M= M_{\rm{c}}$, there is only one horizon: $r_{\rm{h}}=MG_0=\sqrt{\bar{\omega}}$, and there exists $r_*$ satisfying equations (\ref{ineq1}) and $r_*\geq r_{\rm{h}}$; namely, the critical points is outside the outer horizon.

\section{The polytropic solution}
In this section, we only discuss the case where the critical points are outside the outer horizon of the improved Schwarzschild hole. We consider quantum corrections to the temperature and the observed total integrated flux which might be observed. Following Refs. \cite{michel1972accretion}, one can introduce the polytrope equation of state
\begin{eqnarray}
\label{eos}
p=K\rho^\gamma,
\end{eqnarray}
where the adiabatic index $\gamma$ satisfies $1<\gamma<5/3$ and $K$ is a constant. The temperature $T$ of the gas can be obtained from the equation of state of the ideal gas, $p=k_{\rm B}\rho T/(mm_{\rm p})$, where $k_{\rm B}$ is the Boltzmann constant, $m$ is the mean molecular weight, and $m_{\rm p}$ is the mass of the proton. The density and the pressure can be expressed as functions of temperature, respectively
\begin{eqnarray}
\label{eos1}
\rho=\frac{1}{K^n}T^n_{\rm p},~~~~~p=\frac{1}{K^n}T^{n+1}_{\rm p},
\end{eqnarray}
where $n=1/(\gamma-1)$ and $T_{\rm p}=k_{\rm B} T/(mm_{\rm p})$. For constant $\gamma$, we have $\varepsilon+p=(n+1)p$ \citep{michel1972accretion}. Using the equation of state of the ideal gas, equation (\ref{V}) can be reexpressed as
\begin{eqnarray}
\label{V1}
V^2=\frac{d\ln(\mu+p)}{d\ln \rho}-1=\frac{(n+1)T_{\rm p}}{n[1+(n+1)T_{\rm p}]}.
\end{eqnarray}
Using this equation, the improved Bernoulli equation (\ref{con3}) can be rewritten as
\begin{eqnarray}
\label{con6}
\left[1+(n+1)T_{\rm p}\right]^2\left[1-\frac{2MG(r)}{r}+u^2\right]=\left[1+(n+1)T_{\rm p\infty}\right]^2=C_3.
\end{eqnarray}
At infinity we expect that $T_{\rm p\infty}$ is very small and $C_3$ is nearly unity
\begin{eqnarray}
C_3\simeq 1+2(n+1)T_{\rm p\infty}.
\end{eqnarray}
At the critical points, from equations (\ref{critic1}) and (\ref{V1}), we obtain
\begin{eqnarray}
T_*=\frac{nu^2_*}{(n+1)(1-4u^2_*-2MG')}\simeq\frac{n}{n+1}u^2_*.
\end{eqnarray}
Furthermore, equations (\ref{critic1}), (\ref{V1}), and  (\ref{con6}), lead to
\begin{eqnarray}
C_3\simeq \frac{[1-(n-4)u^2_*]^2(1-3u^2_*-2MG')}{(1-4u^2_*-2MG')^2}\simeq 1+(2n-3)u^2_*-2MG'(r_*).
\end{eqnarray}
Then we have
\begin{eqnarray}
T_*=\frac{2n}{2n-3}T_{\rm p\infty}+\frac{2n}{(n+1)(2n-3)}MG'(r_*).
\end{eqnarray}
From equations (\ref{con1s1}) and (\ref{eos1}), it is easy to get
\begin{eqnarray}
T^n_{\rm p}ur^2=C_4,
\end{eqnarray}
where $C_4$ is a constant. So
\begin{eqnarray}
\label{c4}
C_4=T^n_*u_*r^2_*=\frac{(2n)^{n}G^{2}_{0} M^2 T^{n-3/2}_{\rm p\infty}}{4[2(n+1)]^{\frac{3}{2}}(2n-3)^{n-\frac{3}{2}}}\left[1+\frac{2n-3}{2(n+1)}\frac{MG'(r_*)}{T_{\rm p\infty}}\right],
\end{eqnarray}
which means that all the constants are determined in terms of $T_{\rm p\infty}$. As a first approximation, we assume that $T_{\rm p\infty}$ is much less than unity, equation (\ref{con6}) immediately gives $u(r)\simeq \sqrt{2MG(r)/r}$. Therefore from equation (\ref{c4}), we derive the temperature
and the matter density as, respectively,
\begin{eqnarray}
T_{\rm p}=\frac{\sqrt{2}n T^{1-3/2n}_{\rm p \infty}}{4[2(n+1)]^{\frac{3}{2n}}(2n-3)^{1-\frac{3}{2n}}}\left(\frac{MG_0}{r}\right)^{\frac{3}{2n}}\left[1+\frac{2n-3}{2n(n+1)}\frac{MG'(r_*)}{T_{\rm p\infty}}\right],
\end{eqnarray}
\begin{eqnarray}
\rho(r)=\left(\frac{\sqrt{2}n}{4 K}\right)^n\frac{T^{n-3/2}_{\rm p\infty}}{[2(n+1)]^{\frac{3}{2}}(2n-3)^{n-\frac{3}{2}}}\left(\frac{MG_0}{r}\right)^{\frac{3}{2}}\left[1+\frac{2n-3}{2(n+1)}\frac{MG'(r_*)}{T_{\rm p\infty}}\right],
\end{eqnarray}
where the constant $K$ is unknown. In other words, the matter density does not appear explicitly, but can be given from
\begin{eqnarray}
\frac{\rho(r)}{\rho_\infty}=\left(\frac{T_{\rm p}}{T_{\rm p\infty}}\right)^n.
\end{eqnarray}
The improved Bondi accretion rate can be determined by using the critical point
\begin{eqnarray}
\label{acc1}
\dot{M}=4\pi r^2_*u_*\rho_*=\pi\left(\frac{2n}{K}\right)^n\frac{G^{2}_{0}M^2T^{n-3/2}_{\rm p\infty}}{[2(n+1)]^{\frac{3}{2}}(2n-3)^{n-\frac{3}{2}}}\left[1+\frac{2n-3}{2(n+1)}\frac{MG'(r_*)}{T_{\rm p\infty}}\right],
\end{eqnarray}
substituting this equation into (\ref{intflux}), we can easily get the improved observed total integrated flux. Compared with the normal total integrated flux $F_{\nu0}$, the extra integrated flux resulting from quantum gravity is
\begin{eqnarray}
\label{extraintflux}
\digamma=\frac{\triangle F_\nu}{F_{\nu0}}=\frac{F_\nu-F_{\nu0}}{F_{\nu0}}=\frac{2n-3}{2(n+1)}\frac{MG'(r_*)}{T_{\rm p\infty}}.
\end{eqnarray}
At the event horizon of the black hole, $r\simeq 2MG_0$, the temperature $T_{\rm ph}$ of the gas takes the form
\begin{eqnarray}
T_{\rm ph}=\frac{\sqrt{2}n T^{1-3/2n}_{\rm p\infty}}{4[4(n+1)]^{3/2n}(2n-3)^{1-3/2n}}\left[1+\frac{2n-3}{2n(n+1)}\frac{MG'(r_*)}{T_{\rm p\infty}}\right].
\end{eqnarray}
For $\gamma=4/3$ ($n=3$), the extra integrated flux, the temperature $T$ of the gas at horizon and at critical point take values, respectively
\begin{eqnarray}
\label{extraintflux}
\digamma=\frac{3}{8}\frac{MG'(r_*)}{T_{\rm p\infty}},
\end{eqnarray}
\begin{eqnarray}
T_{\rm ph}=\frac{\sqrt{6}}{16}T^{1/2}_{\rm p\infty}\left[1+\frac{1}{8}\frac{MG'(r_*)}{T_{\rm p\infty}}\right],
\end{eqnarray}
and
\begin{eqnarray}
T_*=2T_{\rm p\infty}+\frac{1}{2}MG'(r_*).
\end{eqnarray}
For $r\gg \sqrt{\bar{\omega}}$ or equally $M\gg m_{\rm p}$, we have $\digamma\sim 0$ and the quantum gravity corrections to $T_{\rm ph}$ is also nearly zero, see for example, taking $T_{\rm p\infty}\sim10^{-9}$, $M\sim 3M_\odot$, and $r_*\sim 3R_\odot$, we get $\digamma\sim \mathcal{T}_{\rm ph} \sim 10^{-68}$ where $\mathcal{T}_{\rm ph}\equiv (T_{\rm ph}-T_{\rm ph0})/T_{\rm ph0}$ and $T_{\rm ph0}$ the quantity when quantum gravity effects are not taken into accounted. Only for $r\sim \sqrt{\bar{\omega}}$, the quantum gravity corrections to $F_\nu$ and $T_{\rm ph}$ are comparable; see for example, taking $T_{\rm p\infty}\sim10^{-9}$, $r_*\sim 10\sqrt{\bar{\omega}}$, and $M\sim 5M_{\rm c}$, we have $\digamma\sim \mathcal{T}_{\rm ph} \sim 10^6$. Quantum gravity correction to $T_*$ is very small, even near the Planck scale.

We note, however, here we do not take into account the back-reaction of accreting perfect fluid. To discuss the back-reaction in a full way is notoriously difficult, analytic solutions are discussed in \citep{Tolman:1934za, Vaidya:1951zz, Vaidya:1953zza, Vaidya:1999zz, Wang:1998qx, Malec:1999dd, Karkowski:2005zn}. For complicated cases, perturbative scheme is often adopted \citep{Dokuchaev:2011gt, Babichev:2012sg}. When the black hole is very large, the back-reaction can be neglected, but if the mass of the black hole reduces to a certain critical value or the mass of the accreting gas exceeds a certain critical value, the the back-reaction should be taken into account, for details see \citep{Malec:1999dd, Karkowski:2005zn}. In other words, the self-gravity effects of the in-falling flow might be important \citep{Malec:1999dd, Karkowski:2005zn}. However, if back-reaction of the fluid are considered, we can still treat the quantum gravity effects as the corrections to the total effects of the accretion and the back-reaction. This problem worth deeply discussing but is very complicated, it is beyond the scope of this paper, and we leave it for future investigation.

\section{Conclusions and discussions}
In recent years, asymptotic safety conjure has attracted much attentions and some interesting results have been found, see, for example, quantum gravity corrections will change the structure of the Schwarzschild black hole \citep{Bonanno:2000ep}. These changes will affect the physical prosse of the black hole, such as the black hole thermodynamics.

Here we investigated effects of quantum gravity to accretion onto black holes in the context of asymptotically safe scenario. We considered steady, spherical accretion onto a static and spherically symmetric improved Schwarzschild black hole. We discussed the possible positions of the critical points and obtained the general conditions for critical points. We derived the analytic general expressions for the critical fluid velocity. We also found the explicit expressions for matter density compression and temperature profile both below the critical radius and at the event horizon. For polytropic matter, we determined the corrected temperature and the integrated flux resulting from quantum gravity effects at the event horizon, which might be as a test of asymptotically safe scenario.

\begin{acknowledgments}
I thank Anslyn J. John for helpful discussions. This study is supported in part by National Natural Science Foundation of China (Grant Nos. 11147028 and 11273010), the Hebei Provincial Outstanding Youth Fund (Grant No. A2014201068), the Outstanding Youth Fund of Hebei University (No. 2012JQ02), the Open Project Program of State Key Laboratory of Theoretical Physics, Institute of Theoretical Physics, Chinese Academy of Sciences, China (No.Y4KF101CJ1), and the Midwest universities comprehensive strength promotion project.
\end{acknowledgments}

\bibliography{ref}

\end{document}